\documentclass{SciPost}

\hypersetup{
    colorlinks,
    linkcolor={red!50!black},
    citecolor={blue!50!black},
    urlcolor={blue!80!black}
}

\usepackage[bitstream-charter]{mathdesign}
\DeclareSymbolFont{usualmathcal}{OMS}{cmsy}{m}{n}
\DeclareSymbolFontAlphabet{\mathcal}{usualmathcal}

\fancypagestyle{SPstyle}{
\fancyhf{}
\lhead{\colorbox{scipostblue}{\bf \color{white} ~SciPost Physics Codebases }}
\rhead{{\bf \color{scipostdeepblue} ~Submission }}

\fancyfoot[C]{\textbf{\thepage}}
}

\usepackage{threeparttable}
\usepackage{hyperref}
\usepackage{braket}
\usepackage[normalem]{ulem} 

\newcommand{\autohf}{\texttt{autohf}} 

\usepackage{float}
\usepackage{listings}
\usepackage{bm}

\newfloat{lstfloat}{htbp}{lop}
\floatname{lstfloat}{Listing}

\usepackage[frozencache,cachedir=minted-cache]{minted}
\usepackage{xcolor}
\definecolor{LightGray}{gray}{0.9}

\usepackage{bookmark}

\begin{document}

\pagestyle{SPstyle}

\begin{center}{\Large \textbf{\color{scipostdeepblue}{
AutoHF: a general Hartree-Fock solver utilizing direct energy minimization with automatic differentiation
}}}\end{center}

\begin{center}\textbf{
Ryan Levy\textsuperscript{1$\star$},
Brandon Eskridge\textsuperscript{1$\dagger$},
Lukas Weber\textsuperscript{1$\ddagger$},
Miguel A. Morales\textsuperscript{1}$\mathsection$, and
Shiwei Zhang\textsuperscript{1$\mathparagraph$}
}\end{center}

\begin{center}

{\bf 1} Center for Computational Quantum Physics, Flatiron Institute, New York, NY, 10010, USA
\\[\baselineskip]
$\star$ \href{mailto:email1}{\small ryan@ryanlevyphys.com}\,,\quad
$\dagger$ \href{mailto:keskridge@flatironinstitute.org}{\small keskridge@flatironinstitute.org}
$\ddagger$ \href{lweber@flatironinstitute.org}{\small lweber@flatironinstitute.org}\,,\quad
$\mathsection$ \href{mailto:mmorales@flatironinstitute.org}{\small mmorales@flatironinstitute.org}\,,\quad
$\mathparagraph$ \href{mailto:szhang@flatironinstitute.org}{\small szhang@flatironinstitute.org}
\end{center}

\section*{\color{scipostdeepblue}{Abstract}}
\textbf{\boldmath{%
We present \autohf{}, a general, easy-to-use mean-field solver for quantum many-fermion Hamiltonians. 
It allows the user to bypass the process of deciphering the mean-field form for each many-body Hamiltonian $H$ and thus avoid setting up a tailored program for each $H$. Rather, \autohf{} finds the optimal Slater determinant $|\Psi\rangle$, written in terms of orbital coefficients and subject to symmetry constraints, by directly minimizing the variational energy $\braket{H}$. By embracing this variational approach, \autohf{} makes use of the growing power of automatic differentiation and optimization tools developed by the machine learning community.
}}

\vspace{\baselineskip}



\vspace{10pt}
\noindent\rule{\textwidth}{1pt}
\tableofcontents
\noindent\rule{\textwidth}{1pt}
\vspace{10pt}

\section{Introduction}
\label{sec:intro}

The interacting many-electron problem is among the most fundamental and
persistent challenges in modern physics and chemistry.
The non-trivial interplay between one-body effects and electron correlation
underpins phenomena ranging from the Mott transition to
fractional Chern insulators in moir\'e semiconductors~\cite{Cai2023,Lu2024},
magnetic, charge, and superconducting orders, and
the long-standing puzzle of stripe and pseudogap physics in the cuprates~\cite{Qin2020}.
Despite decades of theoretical and computational effort, the exponential scaling of the
many-body Hilbert space with system size renders an exact solution
intractable for all but the smallest systems
or special cases.
A hierarchy of approximate methods are used instead.
A central practical reality, common to applications across a wide range of problems, is the
need for efficient, accurate, and controllable approximate solutions that
can serve both as standalone methods and as starting points for more
accurate many-body treatments such as 
coupled cluster theory~\cite{Bartlett2007},
neural quantum state variational
methods~\cite{Carleo2017,Lange2024},
green's function or diffusion Monte Carlo~\cite{sorella-book,Foulkes2001},
or auxiliary-field quantum Monte Carlo (AFQMC)~\cite{Zhang1997,ZhangKrakauer2003,MottaZhang2018}.

The Hartree-Fock (HF) method is ubiquitous in quantum many-electron theory across a wide range of application domains, including lattice models, moiré systems, quantum chemistry, and 
\textit{ab initio} studies of solids. 
As the simplest variational approach that 
exactly includes exchange, and accounts for interaction effects via self-consistent mean-field theory,
HF occupies a central role in both conceptual understanding and practical interacting electron calculations.
The intrinsic limitations of HF theory itself are well known.
Essentially by definition, HF neglects electron correlation, 
which leads to quantitatively inaccurate results.
In some cases, HF makes even qualitatively incorrect predictions.
An example is the two-dimensional Fermi-Hubbard model where spin-unrestricted HF (UHF), which allows for spin-dependent HF orbitals, agrees with accurate methods across much of the magnetic phase diagram, especially with some rescaling or renormalization of the interaction strength~\cite{Xu2022}, but for
interaction strengths near phase transitions, UHF predicts an incorrect phase~\cite{Xu2011,Qin2016,Zheng2017}.
Nevertheless, HF remains a valuable and widely used tool, both for stand-alone theoretical investigation and as a starting point for more advanced many-body methods. 

The HF ansatz consists of a single Slater determinant and 
the variational freedom is provided by the choice of single-particle orbitals
used to construct the Slater determinant.
In practice, the HF orbitals are most commonly determined using the self-consistent field (SCF) approach.
The single-reference character of the HF state allows an effective mean-field Hamiltonian to be 
constructed analytically, i.e. via Wick's theorem, based on some initial guess for the HF state.
The mean-field Hamiltonian can be solved exactly to arrive at a new guess for the HF state.
SCF arrives at the HF solution by iterating until the HF state converges to within some numerical threshold.
The SCF approach has historically had several important advantages.
The specific form of mean-field decoupling can be used to rigorously preserve or break some symmetries as desired.
In addition, SCF implementations are relatively fast and have low-order polynomial scaling in the system size.
This is important, especially as increasingly large system sizes are treated both in electronic structure of real materials and in lattice models. The importance of computational efficiency is further magnified when multiple HF 
calculations with differing starting points must be performed to ensure that the global
minimum is found.

Many open-source, mature implementations of HF exist for quantum chemistry 
~\cite{PySCF2020,PySCF2018,libcint2015,NWChem2020,GAMESS2020,CP2K_2020},
as well as for electronic structure~\cite{QE2017,QE2009,ABINIT2020,OCTOPUS2020}.
For lattice models, research codes aimed at specific models are often written 
and used on a per-project basis leading to significant code reproduction
and requiring careful testing of new implementations.
Implementing an extensible and maintainable HF code 
is challenging in practice due to both the wide variety of possible 
lattice geometries, Hamiltonian terms, specific mean-field decouplings,
and optimization algorithms that would need to be implemented to cover general lattice models.

An alternative solution strategy to SCF
is to perform direct variational optimization.
In this case, a cost function is defined based on the interacting Hamiltonian,
and the cost function is minimized based on a single-reference ansatz.
Symmetries can be enforced directly in the specific form of a variational ansatz,
as opposed to the mean-field decoupling, or by imposing energetic 
penalties within the cost function.
Recent advances in optimization algorithms and software libraries have made 
this strategy feasible for general Hamiltonians and ansätze.
Automatic differentiation enables derivatives of complex cost functions to be generated with minimal effort, 
allowing state-of-the-art optimizers to be applied to new problems without extensive manual derivations or code development.
Indeed, direct variational optimization HF has already been explored in
the context of quantum chemistry~\cite{Yoshikawa2022,Helal2024,TamayoMendoza2018, Kasim2022, PennylaneChem}, 
and in some cases demonstrated favorable convergence
compared with SCF implementations although at a higher computational cost~\cite{Yoshikawa2022}.

Here we 
present a new software tool, \autohf{}\footnote{\autohf{} Github Repository: \url{https://github.com/SFQMC/AutoHF}}, which is a package that leverages these developments
in automatic differentiation and optimization tools, most notably driven by rapid advances in connection with machine learning,
to provide a general, flexible, and maintainable framework for direct variational Hartree-Fock optimization.
It is implemented entirely in Python using automatic differentiation provided by JAX \cite{jax2018github}, 
and is designed to support rapid experimentation with new models, Hamiltonians, and ansätze.
\autohf{} is able to take advantage of JAX's GPU backend in order to greatly accelerate HF calculations 
on large lattices.
While the primary focus of \autohf{} is on general lattice models, 
this flexibility allows it to handle other classes of systems as we will demonstrate.

In the remainder of this paper, we describe the underlying approach used in \autohf{}, 
provide guidance on using the software, and demonstrate its capabilities through a series of representative examples, structured
as follows. 
In Sec.~\ref{sec:method}, we describe the general formulation of the variational optimization problem. 
In Sec.~\ref{sec:user_guide}, we provide a practical guide to using the software. 
In Sec.~\ref{sec:examples}, we present several worked examples illustrating the flexibility and robustness of the approach. 
Finally, in Sec.~\ref{sec:conclusion} we summarize and discuss possible future directions.

\section{Theory}
\label{sec:method}

We consider interacting $N$-fermion systems 
consisting of $d$ single-particle degrees of freedom, encompassing orbital/band, 
spin, sublattice, and any other quantum numbers inherent to the system at hand. 
The system is described by an interacting Hamiltonian of the form, 
\begin{equation}
\label{eq:simple_general_hamiltonian}
\hat{H} = \hat{T} + \hat{H}^\text{Int},
\end{equation}
where $\hat{T}$ is a non-interacting Hamiltonian, 
and $\hat{H}^\text{Int}$ is the interacting part of the Hamiltonian which 
consists of one or more interaction terms. 
$\hat{T}$ is of bilinear form of fermion creation and annihilation operators on the 
single-particle basis, defined by a basis index $\mu\in[1,d]$. The two-body interaction
$\hat{H}^\text{Int}$ consists of quartic terms with two pairs of creation and annihilation operators and corresponding matrix elements (in Secs.~\ref{sec:interacting_hamiltonians} and \ref{sec:examples} we provide
concrete examples including the
onsite Hubbard $U$, density-density and spin-spin
interactions, Hund's coupling, and the full Coulomb interaction
in a molecular basis).
In the discussion that follows, we focus on the many-electron problem, but 
generalizations to other fermionic species are straight-forward.

HF represents perhaps the simplest approximate solution to the interacting
electron problem.
It is fundamentally a variational approach in which the ansatz consists of a single Slater determinant (SD) $\ket{\Phi}$,
\begin{equation}
\label{eq:slater_determinant}
\ket{\Phi} = \frac{1}{\sqrt{N}}  \begin{vmatrix}
    \Phi_{11} & \Phi_{12} & \Phi_{13} & \cdots & \Phi_{1N} \\
    \Phi_{21} & \Phi_{22} & \Phi_{23} & \cdots & \Phi_{1N} \\
    \vdots & \vdots & \vdots & \ddots & \vdots \\
    \Phi_{d1} & \Phi_{d2}  & \Phi_{d3}  & \cdots & \Phi_{dN} \\
    \end{vmatrix},
\end{equation}
 where the columns, $p$, 
represent single-particle orbitals, $\phi_p$, and rows, $\mu$, correspond to basis functions.
Computationally, the Slater matrix above, $\Phi_{\mu p}$, is often represented by a rectangular array of floating point numbers.

The SD ansatz is typically factorized according to the desired spin symmetry, giving rise to three standard variants of HF, spin-restricted (RHF), spin-unrestricted (UHF), and generalized (GHF), built respectively on \emph{closed}, \emph{collinear}, and \emph{noncollinear} SDs. 
These form a strict hierarchy: every RHF solution is a special case of UHF, and every UHF solution a special case of GHF.
We describe each in order of increasing generality.

In RHF, a closed SD takes the form
\begin{equation}
\label{eq:closed_sd}
| \Phi \rangle = | \Phi_{\uparrow} \rangle \otimes  | \Phi_{\downarrow = \uparrow} \rangle,
\end{equation}
where $\downarrow = \uparrow$ indicates that the spin-down orbitals are
identical to those of the spin-up channel. 
A spin-polarized state is possible and is referred to as restricted open-shell HF (ROHF) in which $\Phi_{\downarrow} $ is 
equal to the first $N_\downarrow$ columns of 
$\Phi_{\uparrow} $ (assuming $N_\downarrow\le N_\uparrow$).
RHF and ROHF preserve spin symmetry, yielding eigenstates of 
$ \hat{S}^z $ and $ \hat{S}^2 $, but may break translation and other symmetries.

UHF uses a collinear SD with $N_{\uparrow}$ spin-up and $N_{\downarrow}$
spin-down electrons,
\begin{equation}
\label{eq:collinear_sd}
| \Phi \rangle = | \Phi_{\uparrow} \rangle \otimes  | \Phi_{\downarrow} \rangle,
\end{equation}
where each spin channel has its own set of HF orbitals $\{\phi_{i\sigma}\}$,
and $\sigma$ is an explicit spin index. Allowing the two channels to vary
independently makes the ansatz more expressive. However, this comes at the cost that the ansatz is not an eigenstate of 
$\hat{S}^2$ in general. They remain, however, rigorous eigenstates of $\hat{S}^z$ with
eigenvalue $\tfrac{1}{2}(N_{\uparrow} - N_{\downarrow})$, since
$N_{\uparrow}$ and $N_{\downarrow}$ are individually conserved.

GHF uses a noncollinear SD, in which each column of the Slater matrix is
an arbitrary spin orbital.
The Slater matrix takes the block form
\begin{equation}
\label{eq:noncollinear_sd}
\Phi_{\mu p} = \begin{bmatrix}
    \Phi_{i\uparrow,p} \\
    \Phi_{i\downarrow,p} \\
    \end{bmatrix},
\end{equation}
where $\mu = (i,\sigma)$ combines orbital and spin row indices, and
$p$ runs over all electrons regardless of spin polarization.
Because only the total electron number $N = N_{\uparrow} + N_{\downarrow}$
is fixed, and not $N_{\uparrow}$ and $N_{\downarrow}$ individually,
a noncollinear SD is not in general an eigenstate of either $\hat{S}^z$ or
$\hat{S}^2$.

The choice of HF ansatz to apply in a specific problem depends on the physics that one wishes to capture. 
For example, GHF is necessary to capture spin-flip effects if spin-orbit coupling is included in the Hamiltonian. 
Even in a situation where there is no spin-orbit coupling, a GHF ansatz can still be used to break symmetry in directions other than $S^z$. 

The HF solution for a given interacting Hamiltonian $H$ is then the SD $\Phi_{\mu p}$ %
that minimizes the variational energy
\begin{equation}
\label{eq:evar_definition}
E^\text{var} = \frac{\langle \Phi | \hat{H} | \Phi \rangle }{\langle \Phi | \Phi \rangle}.
\end{equation}
To obtain this solution, one typically either constructs an explicit effective mean-field Hamiltonian, or 
minimizes the variational energy of the SD ansatz,
possibly under explicit constraints.
As mentioned, the former has been the standard approach taken in most HF calculations to date. In the paper we adopt the latter.

Both approaches ultimately lead 
to the HF equation,
\begin{equation}
\label{eq:hf_equation}
\hat{f}( \Phi ) \phi_p = \epsilon_p \phi_p,
\end{equation}
where $\epsilon_p$ are the eigenvalues / one-body energies of the HF orbitals $\{  \phi_p \}$,
and $\hat{f}(\Phi)$ is the Fock operator given by,
\begin{equation}
\label{eq:fock_operator}
\hat{f}( \Phi ) = \hat{T} + \hat{v}^\text{HF}( \Phi ),
\end{equation}
and $\hat{v}^\text{HF}( \Phi )$ is the HF potential which is a mean-field term 
that captures the interaction between each electron and the 
average density of the remaining electrons.
The explicit form of $\hat{v}^\text{HF}( \Phi )$ depends on the specific interactions
included in the Hamiltonian.
We leave this form generic since, as we will see below, direct variational optimization
formulation 
does not require an explicit mean-field potential, $\hat{v}^\text{HF}( \Phi )$.
The Fock operator of equation~\ref{eq:fock_operator} is an effective one-body 
operator which can be diagonalized to obtain a set of orbitals and eigenvalues 
from which the HF Slater determinant can be constructed.
However, it is important to note that $\hat{v}^\text{HF}( \Phi )$ depends explicitly on the
HF orbitals. 
This motivates solving the HF equations self-consistently via the SCF method by constructing some initial guess, diagonalizing $\hat{f}$ to construct a new guess, then iterating until convergence is achieved.

Alternatively to the SCF method, the variational optimization can also be performed directly.
To this end, we define the cost function
\begin{equation}
\label{eq:cost_function}
\xi (\Phi) = E^\text{var} ( \Phi ) + \sum_b \lambda_b C_b( \Phi ), 
\end{equation}
where the
$C_b( \Phi ) $ are optional constraints with penalty weights $\lambda_b$. The constraints can be used to influence the symmetry properties of the resulting HF ground state. We find that adding constraints is often not necessary to arrive at the correct HF ground state.
Even an orthonormality constraint can be excluded in practice.
Of course, the resulting HF orbitals are not orthonormal and, 
in extreme cases, can become nearly linearly dependent. To prevent this, it is typically sufficient to simply orthonormalize the HF orbitals via e.g. a QR decomposition or unitary parameterization.

An observation which is central to the implementation of HF in \autohf{} is the fact that,
using Wick's theorem, 
the variational energy can be mapped to a function of 
the 1-body reduced density matrix (1-rdm), $\rho$ given by 
\begin{equation}
\label{eq:onerdm}
\rho_{\mu\nu} \equiv \frac{\braket{ \Phi |\hat{c}^\dagger_{\mu} \hat{c}_{\nu} | \Phi }}{\braket{ \Phi | \Phi}} = [ \Phi^* ( \Phi^T \Phi^* )^{-1} \Phi^T ]_{\mu\nu},
\end{equation}
which reduces to $[\Phi^*\, \Phi^T ]_{\mu\nu}$ if $\Phi$ is an orthonormal Slater determinant.
The 1-rdm $\rho$ is directly determined by $\Phi$ which in term completely determines 
$\bar{E}^\text{var}$.
In practice, the HF wavefunction may be expressed as,
\begin{equation}
\label{eq:ansatze_phi}
 \ket{\Phi} = \ket{\Phi(\alpha)},
\end{equation}
where $\alpha$ is some set of variational parameters, and $\ket{\Phi(\alpha)}$ is an
explicit function of $\alpha$.
For example, $\alpha$ could be a Slater matrix $\Phi_{\mu p}$.
The 1-rdm can now be expressed as a function of $\alpha$, $\rho(\alpha)$. 
HF is performed by minimizing the cost function $\xi (\Phi(\alpha))$ with respect to $\alpha$.
We note that this framework directly supports 
multi-SD wavefunctions
although multi-SD wavefunctions are beyond the scope of the present work.

\section{User Guide} 
\label{sec:user_guide}

\autohf{} provides a modular interface for defining a Hartree--Fock calculation, selecting a variational ansatz, and minimizing the corresponding HF energy.
A calculation is specified by an \texttt{AutoHFHamiltonian}, 
which collects the kinetic and interaction terms, together with an ansatz that maps variational parameters to a Slater determinant and its 1-rdm. 
Runtime choices are passed through a settings dictionary, including the optimizer backend and any backend-specific options used to carry out the direct energy minimization.
Users may rely on the built-in Hamiltonian terms, ansätze, and energy functions, or supply custom energy terms to handle more general systems.
In the standard workflow, the user constructs the Hamiltonian, chooses the ansatz and settings, and calls \texttt{solve\_hf}.

In \autohf{}, an ansatz consists of the following ingredients.
First, a specific parameterization of the wavefunction, $| \Phi(\alpha) \rangle$ is needed, where $\alpha$ is some set of parameters. 
For most of the built-in ansätze (see Section~\ref{sec:variational_ansatze}), $\alpha$ is one or more matrices.
In practice, this must be implemented as an explicit function which accepts $\alpha$ as input and returns an explicit Slater matrix which
consists of an explicit set of HF orbitals.
Second, an ansatz requires a function for computing the 1-rdm from $\alpha$, $\rho(\alpha)$.
This is equivalent to computing $\rho$ as a functional of $|\Phi(\alpha) \rangle$, but it allows the 1-rdm to be partially evaluated analytically
for each specific ansatz.

Since an \autohf{} ansatz maps $\alpha$ to $\rho$, the result is a cost function which can be directly optimized in the parameters $\alpha$
using existing optimizers.
\autohf{} currently interfaces to several different optimization methods implemented in existing libraries. 
We currently interface to gradient descent, as implemented in the Optax Python library, 
LBFGS~\cite{LBFGS_original} as implemented in the jaxopt Python library, 
and the basin hopping algorithm~\cite{Basinghopping_original} as implemented in the scipy.optimize library.

Since the optimization problem is comparatively small and not noisy 
we find that LBFGS generally outperforms first-order optimizers such 
as Adam that are popular in large scale ML applications. 
We also did not find a significant benefit from basin hopping, 
compared to running an ensemble of optimizations from different starting points in parallel.
For this reason, LBFGS is the default optimizer in \autohf{}.
Interfacing to additional optimization algorithms is straight forward and simply requires setting up hand shaking 
between the \autohf{} (usually JAX) types and the library where the new algorithm is implemented.
\autohf{} batches parallel HF trials when invoking the optimizers since many trials are necessary in general in 
order to reliably find the best HF solution.

Below we describe how to run \autohf{} in section~\ref{sec:running_autohf}, 
along with an \autohf{} settings reference,
and a reference for passing settings to the optimizers on the backend.
We then explain how to set up a Hamiltonian in section~\ref{sec:interacting_hamiltonians},
and the implemented ansätze in section~\ref{sec:variational_ansatze}.

\subsection{Running \autohf{}}
\label{sec:running_autohf}

\autohf{} is invoked from within a Python script by calling its main function, \texttt{solve\_hf}.
At a minimum, \texttt{solve\_hf} requires an \texttt{AutoHFHamiltonian} and a Python dictionary 
containing settings as input.

\begin{lstfloat}
\begin{minted}
[
frame=lines,
framesep=2mm,
baselinestretch=1.2,
fontsize=\footnotesize,
linenos
]{python}
# File: run_autohf.py
# 
# how to run a Python script: $ python run_autohf.py from a terminal
import autohf

# 1. setup the Hamiltonian
Nx, Ny = 4, 4
T = makeT(Nx, Ny) # generates the hopping matrix - definition shown below
H = autohf.AutoHFHamiltonian(
    T=(T, T), 
    U=2.0
)

# 2. Choose settings
hf_settings = {
  "steps": 100,
  "ansatz": "SD", 
  "batch_size": 20,
  "nelec": (8, 8),
  "noncollinear" : False
}

# 3. invoke autohf
results = autohf.solve_hf(
  H,
  settings=hf_settings,
)
\end{minted}
\caption{Example of running autohf. The \texttt{makeT} function is defined in listing~\ref{lst:ex_makeT_square_lattice}.}
\label{lst:running_autohf}
\end{lstfloat}
Listing~\ref{lst:running_autohf} contains an example of running \autohf{} for the Hubbard model (specifically a $4 \times 4$ square lattice with periodic boundary conditions and $U=2$).
First, the \texttt{AutoHFHamiltonian} is created by providing it with a value for the onsite Hubbard interaction, 
and the hopping matrix as a tuple of arrays, \mintinline{python}{(T,T)}, where the elements
of the tuple correspond to the spin-up, and spin-down channels respectively, and \textit{T} is an array with shape $(N_\text{sites},N_\text{sites})$.
Next, we define a Python dictionary containing some settings for \autohf{}.
The settings exposed in this example are some of the most common settings that a typical user will interact with.
For example, the parameterization of the HF ansatz is chosen using the \texttt{"ansatz"} keyword.
In this case, we are using the \texttt{"SD"} ansatz which corresponds to directly optimizing the Slater matrix.
We will explain the implemented ansätze in detail in section~\ref{sec:variational_ansatze}.
The number of electrons, \texttt{"nelec"}, is set as a tuple, $(N_{\uparrow},N_{\downarrow})$, 
the maximum number of optimization steps to take is provided using \texttt{"steps"}, 
and the \texttt{"batch\_size"} sets the number of HF trials to run in parallel.
The calculation will stop when all trials have converged, or when the maximum 
number of steps has been reached.
Finally, the keyword \texttt{"noncollinear"} turns on/off spin noncollinear mode, i.e. GHF.
UHF is the default and RHF/ROHF can be run as special cases of UHF using a custom
ansatz as shown in section~\ref{sec:custom_ansatz}.

\begin{lstfloat}
\begin{minted}
[
frame=lines,
framesep=2mm,
baselinestretch=1.2,
fontsize=\footnotesize,
linenos
]{text}

 ===== AutoHF Settings ===== 

        noncollinear                                            False
        gpu                                                     False
        approx_expm                                             False
        force_complex                                           False
        ansatz                                                     SD
        opt_method                                              lbfgs
        riemannian                                              False
        seed                                               1767199325
        eta                                                       0.1
        steps                                                     100
        batch_size                                                 20
        plot                                                    False
        verbose                                                 False
        output                                                   None
        dump_batch                                              False
        measure_spin                                            False
        random_initial_phi0                                     False
        state0_scale                                             0.01
adding flags
[CpuDevice(id=0)]
default dtype: float64
nelec: (8, 8)
Using a free electron initial guess
Running in Slater Determinant Mode
Reference HF Energy = -15.464724535019071
Initial Energy: [ 7.08329921  9.0070837   7.26570018  4.42207019  7.33044907  7.19441979
  9.64040134  9.84581397 10.02311618  6.73717983  8.44936551  5.89483673
  7.87302665  7.6366232   8.15118461  7.98348186  7.24390838  5.62211161
  8.71251415 10.87802365]
# iterations: [70 31 62 54 49 48 27 37 47 58 35 73 40 31 29 28 81 66 39 30]
 frac of possible: [0.7  0.31 0.62 0.54 0.49 0.48 0.27 0.37 0.47 0.58 0.35
 0.73 0.4  0.31 0.29 0.28 0.81 0.66 0.39 0.3 ]
Final energy: -17.554570105149665 [-16.99710657 -16.99153534 -16.99710664
 -16.99710691 -16.99710626 -16.99710687 -16.97940524 -16.98820167 -16.98197406
 -16.9971069 -16.99128356 -16.9971069  -16.99706103 -17.55457011 -17.55410902
 -17.55263842 -16.99710685 -16.99710688 -16.99537795 -16.99027177]
Variational energy of final HF state: -17.554570105149665
\end{minted}
\caption{Sample \autohf{} output corresponding to listing~\ref{lst:running_autohf}.}
\label{lst:autohf_output}
\end{lstfloat}

Running the example in listing~\ref{lst:running_autohf} produces the output shown in listing~\ref{lst:autohf_output}.
The \autohf{} settings are echoed at the beginning of the output.
The variational energy of the reference Slater determinant is printed; 
for the \texttt{"SD"} ansatz, the reference Slater determinant can either be provided by the user, or 
a non-interacting guess can be generated internally.
Next the initial energy is printed for each HF trial in the batch.
For the \texttt{"SD"} ansatz, this corresponds to the reference determinant with random noise added
where the amplitude of the noise can be controlled using the \texttt{"state0\_scale"} keyword which defaults to 0.01.
Finally, \autohf{} outputs the number of iterations each trial took to converge, both as an absolute count and as a fraction
of the maximum number of steps.
It is important to check that most trials converged by verifying that the number of actual steps taken is less
than the maximum number of steps.
\autohf{} will print a warning if the best variational energy corresponds to an unconverged trial.
The final energies of each HF trial in the batch are printed, followed by the single best HF energy across all trials.
We take the HF trial with the best variational energy as the HF result.

\label{sec:autohf_settings}

\autohf{} is controlled primarily by passing a dictionary of keyword arguments to its main function.
A few of the most common parameters were demonstrated above.
Here, we provide a reference of the most common settings available in \autohf{}.
The more advanced settings are described in Appendix~\ref{appendix:autohf_settings}.
Table~\ref{tbl:AuotHF_common_settings} lists the most common options along with the default value,
and a description of the setting.

\begin{lstfloat}
\begin{minted}
[
frame=lines,
framesep=2mm,
baselinestretch=1.2,
fontsize=\footnotesize,
linenos
]{python}
hf_settings = {
  "steps": 100,
  "opt_method": "lbfgs",
  "ansatz": "SD", 
  "batch_size": 20,    # default: 1
  "gpu": False,
  "nelec": (8, 8),
  "noncollinear": False,
  "seed": 42           # default: auto-generated seed
}
\end{minted}
\caption{Sample \autohf{} settings dictionary with common parameters exposed. All values displayed are defaults unless otherwise noted.}
\label{lst:autohf_settings_dict}
\end{lstfloat}

\begin{threeparttable}[h]
\begin{tabular}{ p{3cm} p{2cm} p{7cm}  }
 \hline
  \textbf{Parameter} & \textbf{Default} & \textbf{Description} \\
 \hline
  noncollinear & False & if True, enables noncollinear / GHF calculations, \\
  gpu   & False    & If True, enables GPU backend. Requires installing one the optional cuda JAX versions. \\
  ansatz & SD\_ROT & Selects the specific HF ansatz to use. Choices are: SD / SDC for Slater determinant with real / complex valued orbitals, SD\_ROT / SD\_ROTC for rotated Slater determinant  with real / complex valued orbitals, DIAG for a mean-field-like ansatz where orbitals are the eigenvalues of an effective one-body Hamiltonian. CUSTOM if you wish you provide your own custom ansatz \\
  opt\_method & lbfgs & choices are "grad" for gradient decent, "lbfgs" for the L-BGFS algorithm, or "basin" for the basin hopping algorithm. \\
  seed & random & The random seed to use. Will be determined automatically if not provided by user. \\
  steps & 100 & the maximum number of optimization steps to use. \\
  batch\_size & 1 & the number of parallel HF trials to include in a batch. \\
\hline
\end{tabular}
\caption{Common parameters in the \texttt{settings} dictionary passed to the \texttt{solve\_hf} function. For an exhaustive reference see Appendix~\ref{appendix:autohf_settings}.}
\label{tbl:AuotHF_common_settings}
\end{threeparttable}

\subsection{Many-Electron Hamiltonians}
\label{sec:interacting_hamiltonians}

The \autohf{} Hamiltonian can be thought of as
\begin{equation}
\label{eq:general_hamiltonian}
\hat{H} = \ \hat{T} + \hat{H}^\text{Int} = \hat{T} + \sum_{X} \hat{H}^\text{Int}_X,
\end{equation}
where $\hat{H}^\text{Int}$ has been separated into a sum over a set of interaction terms,
$\hat{H}^\text{Int}_X$.
Each Hamiltonian term has a corresponding $\bar{E}_X^\text{var}(\rho)$ which is included in the 
total variational energy, $\bar{E}^\text{var}(\rho)$.
In listing~\ref{lst:ex_simple_hamiltonian},  we demonstrate how to select Hamiltonian terms to include
by providing the corresponding parameters to the \texttt{AutoHFHamiltonian}.
The \texttt{AutoHFHamiltonian} is later used internally to generate $\bar{E}^\text{var}(\rho)$.

\begin{lstfloat}
\begin{minted}
[
frame=lines,
framesep=2mm,
baselinestretch=1.2,
fontsize=\footnotesize,
linenos
]{python}
import autohf

Nx, Ny = 4, 4

T = makeT(Nx, Ny) # definition shown below
H = autohf.AutoHFHamiltonian(
    T=(T, T), 
    U=4.0
)
\end{minted}
\caption{Example of setting up a Hamiltonian. Definition of \mintinline{python}{makeT(Nx,Ny)} is in listing~\ref{lst:ex_makeT_square_lattice}.}
\label{lst:ex_simple_hamiltonian}
\end{lstfloat}

\begin{lstfloat}
\begin{minted}
[
frame=lines,
framesep=2mm,
baselinestretch=1.2,
fontsize=\footnotesize,
linenos
]{python}
import numpy as np

def makeT(Nx, Ny, xperiodic=True, yperiodic=True):
  N = Nx * Ny
  T = np.zeros([N, N])
  # fill hopping matrix
  for i in range(Nx):
    for j in range(Ny):
      a = i * Ny + j
      if i + 1 < Nx or yperiodic:
        b = ((i + 1) % Nx) * Ny + j
        T[a, b] = -1
        T[b, a] = -1
      if i + 1 < Ny or xperiodic:
        b = i * Ny + (j + 1) % Ny
        T[a, b] = -1
        T[b, a] = -1
  return T
\end{minted}
\caption{Example of generating a hopping matrix. \texttt{makeT} generates and returns the nearest-neighbor hopping matrix on a
$N_x \times N_y$ square lattice. 
Periodic boundary conditions are used along both the X,Y axes.}
\label{lst:ex_makeT_square_lattice}
\end{lstfloat}

In the remainder of this section, we list each implemented interaction term, $\hat{H}_X^\text{Int}$, 
provide a brief explanation of what the term is, 
and give a code snippet demonstrating how to add the term to the Hamiltonian.
In addition to the implemented interaction terms,
arbitrary noninteracting Hamiltonians are supported, as described in Section~\ref{sec:noninteracting}, 
as well as user-specified energy terms, as described in Section~\ref{sec:custom_energy_term}.
We note that user-specified energy terms can also be used to impose constraints via an energy penalty.
 
\subsubsection{Noninteracting Term}
\label{sec:noninteracting}

\autohf{} supports general noninteracting Hamiltonians of the form,
\begin{equation}
\label{eq:H_0_general}
\hat{T}  =  \sum_{i \sigma, j \sigma'} T_{i \sigma, j\sigma'}  \hat{c}^\dagger_{i\sigma}\hat{c}_{j\sigma'},
\end{equation}
where $T_{i \sigma, j\sigma'}$ may be complex or real-valued.
The implementation is generic and is agnostic to the details of the underlying system.
This allows an arbitrary model to be handled so long as the user is able to generate 
the corresponding matrix.
An example of constructing the hopping matrix for an $N_x \times N_y$ square lattice can be seen in listing~\ref{lst:ex_makeT_square_lattice}.

When specifying the non-interacting Hamiltonian, \texttt{AutoHFHamiltonian} always expects
the parameter $T$ 
to have shape $( nspin, npol \times M, npol \times M )$, where $nspin$ is the number of 
explicit spin sectors in the basis ( 2 for collinear mode, and 1 for noncollinear mode),
and $npol$ is the number of spin polarizations (1 for collinear and 2 for noncollinear).
For a model with a spin-independent T, the user should specify the same noninteracting Hamiltonian
for both spin sectors when working in collinear mode, as shown in listing~\ref{lst:ex_simple_hamiltonian},
or, if working in noncollinear mode, the user should construct a block-diagonal noninteracting Hamiltonian.
An example of the later case is provided in listing~\ref{lst:closed_to_noncollinear}.

\begin{lstfloat}
\begin{minted}
[
frame=lines,
framesep=2mm,
baselinestretch=1.2,
fontsize=\footnotesize,
linenos
]{python}
import numpy as np
import autohf

Nx, Ny = 4, 4

T = makeT(Nx, Ny) # definition shown below

T_nc = np.block(
  [[T, np.zeros_like(T)],
  [np.zeros_like(T), T]]
)
T_nc =T_nc[None, ... ]

H = autohf.AutoHFHamiltonian(
  T=T_nc, 
  U=4.0
)
\end{minted}
\caption{Example of setting up a noncollinear non-interacting Hamiltonian. The definition of \texttt{makeT(Nx,Ny)} is in listing~\ref{lst:ex_makeT_square_lattice}.
\autohf{} requires a leading dimension with size equal to the number of 
independent spin sectors which, for noncollinear calculations, is one.}
\label{lst:closed_to_noncollinear}
\end{lstfloat}

While we have provided a simple example of a non-interacting Hamiltonian here, very general non-interacting Hamiltonians 
may be used.
For example, including Rashba spin-orbit coupling (SOC) ~\cite{Rosenberg2016},
multi-band models~\cite{Chiciak2018},
quantum chemistry Hamiltonians as seen below in section~\ref{sec:ex_molecules},
and more.

\subsubsection{Hubbard U}
\label{sec:hubbard_u}

\autohf{} supports the standard onsite Hubbard interaction which is given by
\begin{equation}
\label{eq:H_int_hubbard_u}
\hat{H}^\text{Int}_{U} =  \sum_i U_i n_{i\uparrow}n_{i\downarrow},
\end{equation}
where the sum runs over all sites / bands.
Typically, the onsite Hubbard interaction is uniform across all sites.
To reflect this, the \texttt{AutoHFHamiltonian} can accept a single scalar value for $U$
as an input parameter; in this case, 
the simple Hubbard Hamiltonian with on-site interaction $U$ is realized.

Non-uniform Hubbard interactions are also supported and can be included in the 
Hamiltonian by providing a 1-dimensional array, $U[i]$.
The array must have length equal to total size of the basis used, excluding spin degrees of freedom.
For a single-band model, the size of the basis is simply the number of lattice sites. 
In listing~\ref{lst:nonuniform_u} we provide an example of a non-uniform Hubbard interaction
on a $4 \times 4$ square lattice in which $U[i]$ consists of alternating values of $U_a$ / $U_b$.
With the definition of the hopping graph in listing~\ref{lst:ex_makeT_square_lattice}, this example 
corresponds to a Hubbard model with a striped interaction pattern.

\begin{lstfloat}
\begin{minted}
[
frame=lines,
framesep=2mm,
baselinestretch=1.2,
fontsize=\footnotesize,
linenos
]{python}
import numpy as np
import autohf

Nx, Ny = 4, 4
Ua = 1.0
Ub = 2.0
U = np.array([Ua,Ub]*8)

T = makeT(Nx, Ny)
H = autohf.AutoHFHamiltonian(T=(T, T), U=U)
\end{minted}
\caption{Example of setting up a  non-uniform Hubbard interaction. Definition of makeT is in listing~\ref{lst:ex_makeT_square_lattice}}
\label{lst:nonuniform_u}
\end{lstfloat}

\subsubsection{General density-density and spin-spin interactions}
\label{sec:hubbard_u1}

\autohf{} supports interactions with form
\begin{equation}
\label{eq:H_int_hubbard_u1}
\hat{H}^\text{Int}_{U1} =  \sum_{i < j} U_{ij}^1 (\hat{n}_{i\uparrow} \hat{n}_{j\downarrow} + \hat{n}_{i\downarrow} \hat{n}_{j\uparrow} ).
\end{equation}
$\hat{H}^\text{Int}_{U1}$ can be included in the Hamiltonian by supplying the $U_{ij}^1$ array with shape $( M, M )$, where $M$ is the 
size of the basis excluding spin degrees of freedom.
We note that only the upper triangular of $U_{ij}^1$ is actually read and used; all other entries will be ignored by \autohf{}.

\label{sec:hubbard_u2}

\autohf{} also supports interactions with form
\begin{equation}
\label{eq:H_int_hubbard_u2}
\hat{H}^\text{Int}_{U2} =  \sum_{i < j} U_{ij}^2 (\hat{n}_{i\uparrow} \hat{n}_{j\uparrow} + \hat{n}_{i\downarrow} \hat{n}_{j\downarrow} ).
\end{equation}
$\hat{H}^\text{Int}_{U2}$ can be included in the Hamiltonian by supplying the $U_{ij}^2$ array with shape $( M, M )$, where $M$ is the 
size of the basis excluding spin degrees of freedom.
We note that only the upper triangular of $U_{ij}^2$ is actually read and used; all other entries will be ignored by \autohf{}.

With Eqs.~(\ref{eq:H_int_hubbard_u1})
and (\ref{eq:H_int_hubbard_u2}), density-density interactions, $\hat{n}_{i}\,\hat{n}_{j}$, and spin-spin interactions of the form 
$\hat{s}^z_{i}\,\hat{s}^z_{j}$ can be treated.

\subsubsection{Hund's Coupling J}
\label{sec:hubbard_J}

\autohf{} supports Hund's coupling terms with form
\begin{equation}
\label{eq:H_int_hunds_J}
\hat{H}^\text{Int}_J = \sum_{i < j} J_{ij} (\hat{c}^\dagger_{i\uparrow}\hat{c}^\dagger_{j\downarrow}\hat{c}_{i\downarrow}\hat{c}_{j\uparrow}+\hat{c}^\dagger_{i\uparrow}\hat{c}^\dagger_{i\downarrow}\hat{c}_{j\downarrow}\hat{c}_{j\uparrow}+\hat{c}^\dagger_{j\uparrow}\hat{c}^\dagger_{i\downarrow}\hat{c}_{j\downarrow}\hat{c}_{i\uparrow}  +\hat{c}^\dagger_{j\uparrow}\hat{c}^\dagger_{j\downarrow}\hat{c}_{i\downarrow}\hat{c}_{i\uparrow}).
\end{equation}
$\hat{H}^\text{Int}_{J}$ can be included in the Hamiltonian by supplying the $J_{ij}$ array with shape $( M, M )$, where $M$ is the 
size of the basis excluding spin degrees of freedom.
We note that only the upper triangular of $J_{ij}$ is actually read and used; all other entries will be ignored by \autohf{}.

\subsubsection{Cholesky-factorized generic two-body interactions}
\label{sec:cholesky_2body_int}

\autohf{} also supports generic, real-valued two-body interactions provided 
in Cholesky decomposed form~\cite{Beebe1977,Koch2001}.
\begin{equation}
    \hat{H}^\text{Int}_V = \sum_{ijkl} V_{ijkl} \hat{c}^\dagger_{i} \hat{c}^\dagger_{j} \hat{c}_{k} \hat{c}_l =  \frac 12 \sum_{ijkl} \left( \sum_{n=1}^{N_\text{chol}} v^n_{il} v^n_{kj} \right) \hat{c}^\dagger_{i} \hat{c}^\dagger_{j} \hat{c}_{k} \hat{c}_l ,
\end{equation}
where ${v^n_{il}}$ are the so-called Cholesky vectors, and $N_\text{chol}$ is the number of Cholesky vectors.
We provide an example in section~\ref{sec:ex_molecules} where we perform quantum chemistry calculations with this type 
of interaction.

\subsubsection{Custom cost function terms}
\label{sec:custom_energy_term}

\autohf{} supports general, user-supplied cost-function contributions
which are added to $\bar{E}^\text{var} (\rho)$.
These can either represent additional non-interacting terms, interacting terms, or even simply constraints as in equation~\ref{eq:cost_function}.
In listing~\ref{lst:custom_energy_term} we provide a pseudocode example of including a custom energy term.
Custom terms are included by directly providing the corresponding $\bar{E}^\text{var}(\rho)$ to the main function of \autohf{}, \texttt{solve\_hf}.

It is important to note the following requirements for  $\bar{E}^\text{var}(\rho)$ corresponding to a custom energy term.
First, it must be function of the 1-rdm, $\rho$, and not of $\alpha$, which guarantees interoperability with various ansätze 
which handle the mapping from $\alpha$ to $\rho(\alpha)$.
Second, it must be JIT-compilable by JAX. 
We refer the user to the JAX documentation for more details on what Python features are JIT-compilable.
Finally, we provide a concrete example of performing HF using a custom energy term in section~\ref{sec:ex_molecules}.

\begin{lstfloat}
\begin{minted}
[
frame=lines,
framesep=2mm,
baselinestretch=1.2,
fontsize=\footnotesize,
linenos
]{python}
import autohf

def my_custom_energy(state, rdms):
  """
  A custom energy contribution or penalty term
  """
  E = ... # your custom term here
  return E

autohf.solve_hf(
  H,        # any AutoHFHamiltonian
  settings=hf_settings,
  custom_energy_term=my_custom_energy,
)

\end{minted}
\caption{Example of using a custom energy term when running \autohf{}. }
\label{lst:custom_energy_term}
\end{lstfloat}

\subsection{Hartree-Fock Ansätze}
\label{sec:variational_ansatze}

We detail the various implemented wave function ansätze to create HF orbitals for a $M$ site system with $N$ electrons.
The specific ansatz is selected in \autohf{} by settings the \texttt{"ansatz"} keyword to the corresponding value
in the settings dictionary.
While this is written in spirit for a GHF form, by batching two copies of each ansatz we can extend to an explicit UHF form. We include versions for both entirely real and complex modes of parameters. 

\subsubsection{Slater Determinant}

We can directly optimize the orbitals of a Slater determinant as given by the Slater matrix, 
\begin{equation}
    \label{eq:sd_ansatz}
    \Phi_{\mu p}(\alpha) \equiv \mathbf{\alpha}_{\mu p},
\end{equation}
which is represented as a rectangular array (or multidimensional array) of numbers.
This ansatz is selected by setting the \texttt{"ansatz"} keyword to either \texttt{"SD"} or \texttt{"SDC"}.
The latter indicates that complex-valued orbitals will be used while the former corresponds to real-valued orbitals.
The 1-rdm, $\rho(\alpha)$, is given by  equation~\ref{eq:onerdm} as discussed earlier.

Note that a QR decomposition can be added in to ensure orthogonality. 
If this is not done, it is important to check that no near linear dependence has entered into the set of HF orbitals after optimization.
The resulting orbitals should be orthonormalized before use in post-HF methods.

For the explicit Slater determinant ansätze, \autohf{} provides an optional Riemannian optimization mode that can be used to enforce the orbital orthonormality constraint during optimization.
In this mode, the optimization of properly orthonormal Slater determinant orbitals can be thought of as a constrained optimization on a Stiefel manifold $\mathcal{M}$, given by 
$\mathcal{M}=\{\alpha \in \mathbb{R}^{N\times N_e} | \alpha^\dagger \alpha = I_{N_e}\}$.
To bridge between Riemannian optimization and naive unconstrained optimization algorithms,
\autohf{} provides an option to use 
the constraint dissolving (CD) approach~\cite{xiao2022constraint,hu2022constraint}, 
where an alternative objective function $h(\alpha)$ is used instead of the original $\xi(\alpha)$ by setting the \texttt{"riemannian"} \autohf{} setting to \texttt{True} in the settings dictionary.
The new objective function, 
\begin{equation}
    \label{eq:cost_riemannian}
    h(\alpha) = \xi(\mathcal{A}(\alpha)) + \frac \beta 2||c(\alpha)||^2,
\end{equation}
consists of transforming and shifting the parameters, $\alpha$, 
with $\mathcal{A}(\alpha)$ in the argument of the cost function $\xi(\alpha)$ of equation~\ref{eq:cost_function} and applying the constraint as a Lagrange multiplier $c(\alpha)$. The resulting function $h(\alpha)$ rigorously has the same saddle points as the original function, but is no longer restricted to be on the manifold. This allows for a potentially smoother minimization path to the saddle point. For the Slater determinant, $\mathcal{A}(X) = X(3I/2- X^\dagger X/2)$ and $c(X)=||X^\dagger X - I||_F$, and thus requires very little additional overhead to the computation.  
We note that the \texttt{"riemannian"} setting currently 
applies only to the Slater determinant ansätze.

\subsubsection{Unitary Rotation}
Instead of specifying the Slater matrix directly, one can also define the HF state via a unitary rotation
\begin{equation}
\label{eq:sd_rot}
  \ket{\Phi(\alpha)} = e^{X(\alpha)} \ket{\Phi_0},
\end{equation}
from a reference SD state $\ket{\Phi_0}$, where $X$ is a skew-Hermitian matrix ($X = - X^\dagger$).
By Thouless' theorem~\cite{Thouless1960}, this state has a Slater matrix given by
\begin{equation}
\label{eq:sd_rot_thouless}
  \Phi(\alpha)_{\mu p} = [e^{X(\alpha)}]_{\mu \nu} [\Phi_0]_{\nu p},
\end{equation}
where square brackets, $[...]$ indicate the matrix representation.
In practice, we find that even without Riemannian optimization, this ansatz is well behaved for Hubbard-type models.

The 1-rdm, in this ansatz is computed by applying the rotation $e^{X(\alpha)}$ to $\ket{\Phi_0}$ and then using eq.~\ref{eq:onerdm}. Compared to the direct SD ansatz, this comes at a slightly increased numerical cost, but it has the benefit of a more well-behaved parameterization. For example, at $\alpha_i = 0$ the ansatz is exactly $\ket{\Phi_0}$, which should be a reasonable solution and potential local minimum. This is in contrast to the direct SD (\texttt{SD}) where $\alpha_i = 0$ is not defined, leading to a hole in the parameter space.

In \texttt{auto\_hf}, the Unitary Rotation ansatz is selected by setting the \texttt{"ansatz"} keyword to either \texttt{"SD\_ROT"} or \texttt{"SD\_ROTC"}.
The latter indicates that complex-valued orbitals will be used while the former corresponds to real-valued orbitals.
\texttt{"SD\_ROT"} is the default ansatz in \autohf{}.

\subsubsection{Diagonalization}

In spirit to the construction of the self-consistent Hartree-Fock method, 
we can instead define an ansatz as the optimization of orbitals given by a mean-field matrix. 
Let $\tilde{H}_{ij}(\alpha)$ represent a Hermitian matrix with each element of its upper/lower triangular as a parameter,
\begin{equation}
\label{eq:diag_H_tilde}
\tilde{H}_{ij} = T_{ij} + ( \alpha_{ij} + h.c. ),
\end{equation}
where 
$T_{ij}$ is simply the non-interacting part of the Hamiltonian.
We can then obtain the lowest $N_e$ orbitals from the eigenvectors of $\tilde{H}_{ij}$.
The 1-rdm is then constructed from these orbitals in the usual way as per equation~\ref{eq:onerdm}.
The Diagonalization ansatz is selected by setting the \texttt{"ansatz"} keyword to \texttt{"DIAG"}.

\subsubsection{Custom Ansatz}
\label{sec:custom_ansatz}

\autohf{} supports custom ansätze in order allow some constraints to be imposed at the level of the ansatz.
As mentioned above in section~\ref{sec:user_guide}, an \autohf{} ansatz consists of a specific 
parameterization of the wavefunction, $| \Phi(\alpha) \rangle$, in terms of some chosen parameters $\alpha$,
 as well as a function for computing the 1-rdm from $\alpha$, $\rho(\alpha)$.
To use a custom ansätze, both of these functions must be implemented.
For best performance, they should be JIT compiled with JAX as well.

As a simple but illustrative example, 
we implement spin-restricted open-shell Hartree-Fock (ROHF) using a custom ansatz.
In ROHF, the HF orbitals are identical in both the spin-up and spin-down channels, 
but $N_{\uparrow} \neq N_{\downarrow}$ and some orbitals in one spin channel will not be occupied.
We parameterize the HF orbitals as a single Slater matrix, $\Phi_{\mu p} = \alpha_{\mu p}$ with shape $(n_\text{basis},\max(N_{\uparrow},N_{\downarrow}))$ and construct the spin-up (spin-down) 
channels of the Slater determinant using the first $N_{\uparrow}$ ($N_{\downarrow}$) 
orbitals in $\alpha$.
We construct the 1-rdm from the resulting Slater determinant in the usual way.
If the spins of all open-shell electrons are in the same spin channel, then an ROHF Slater determinant is an eigenstate of both $\hat{S}^z$ and $\hat{S}^2$.

Listing~\ref{lst:custom_ansatz_rohf} demonstrates an implementation of the custom ROHF ansatz.
We note that \texttt{orbitalFunc}, which implements $| \Phi(\alpha) \rangle$, requires explicit knowledge of $N_{\uparrow}$ and $N_{\downarrow}$ in order to occupy the orbitals in $\alpha$;
however, \autohf{} expects \texttt{orbitalFunc} to be strictly a function of $\alpha$.
To handle this requirement, \texttt{Nup} ($N_{\uparrow}$) and  \texttt{Ndown} ($N_{\downarrow}$)
are accessed from the surrounding namespace.
JAX's JIT compiler treats them as implicit static arguments.
If the values of \texttt{Nup} and \texttt{Ndown} change, a recompilation of \texttt{orbitalFunc} will be triggered.
It is recommended that metaparameters passed in this way are not changed during the execution of the program.

\begin{lstfloat}
\begin{minted}
[
frame=lines,
framesep=2mm,
baselinestretch=1.2,
fontsize=\footnotesize,
linenos
]{python}
import jax
import jax.numpy as jnp

Nup = 13
Ndown = 12

@jax.jit
def orbitalFunc(alpha):
  r"""
  produces |\Phi_\sigma (alpha) \rangle from \Phi_{\mu p} \equiv \alpha_{\mu p}
  using the first Nup / Ndown electrons for the spin-up / spin-down sectors.
  """
  phi_up = alpha[:,:Nup]
  phi_down = alpha[:,:Ndown]
  return (phi_up, phi_down)
  
@jax.jit
def rdmFunc(alpha):
  r"""takes (M,max(Nup,Ndown)) orbitals
  returns <\psi|c^dag c | \psi>/<\psi|\psi>
  with shape (2,M,M) where first axis is spin
  """
  phi_up,phi_down = orbitalFunc(alpha)
  rdm_up = phi_up.conj() @ jnp.linalg.solve(phi_up.T @ phi_up.conj(), phi_up.T)
  rdm_down = phi_down.conj() @ jnp.linalg.solve(phi_down.T @ phi_down.conj(), phi_down.T)
  return (rdm_up, rdm_down)
\end{minted}
\caption{Example of a custom ansatz. We implement ROHF using spin collinear mode 
and a custom ansatze to enfore that the HF orbitals are identical in both spin channels.
We again use \texttt{makeT} as defined in listing~\ref{lst:ex_makeT_square_lattice}.}
\label{lst:custom_ansatz_rohf}
\end{lstfloat}

\begin{lstfloat}
\begin{minted}
[
frame=lines,
framesep=2mm,
baselinestretch=1.2,
fontsize=\footnotesize,
linenos
]{python}
import jax.numpy as jnp
import numpy as np

import autohf

Ne = max(Nup, Ndown)
Nelec = (Nup,Ndown)

hf_settings = {
  "steps": 500,
  "opt_method": "lbfgs",
  "ansatz": "CUSTOM",
  "batch_size": 200,
  "nelec": (Nup, Ndown)
}

rng = np.random.default_rng(42)

state0_ref = rng.normal(scale=0.2, size=(N, Ne))
state0 = state0_ref + rng.normal(scale=0.01, size=(hf_settings["batch_size"], N, Ne))

autohf.solve_hf(
  H,
  settings=hf_settings,
  state2orbitals=orbitalFunc,
  state2rdm=rdmFunc,
  state0=state0,
  state0_ref=state0_ref,
  jaxoptargs=dict(tol=1e-10), #keyword settings passed to jaxopt
)
\end{minted}
\caption{Example of invoking \autohf{} with a custom ansatz. This is a continuation 
of listing~\ref{lst:custom_ansatz_rohf}.}
\label{lst:run_custom_ansatz}
\end{lstfloat}

\begin{lstfloat}
\begin{minted}
[
frame=lines,
framesep=2mm,
baselinestretch=1.2,
fontsize=\footnotesize,
linenos
]{python}
# N, Ne defined elsewhere
# also reference hopping matrix TGHF_ which is (2*N,2*N)
signs = (-1)**np.arange(N)

def state2orbitals(state):
    # take the hopping matrix TGHF, and "set" M on the off-diagonal blocks
    TGHF = TGHF_+0.0
    TGHF = TGHF.at[N+np.arange(N),np.arange(N)].set(signs*state)
    TGHF = TGHF.at[np.arange(N),N+np.arange(N)].set(signs*state)
    lams,vecs = jnp.linalg.eigh(TGHF)
    return jnp.stack([vecs[:,:2*Ne]])
    
def state2rdm(state):
    orbitals = state2orbitals(state)
    return autohf.tools.sd_to_rdm_noncollinear(orbitals)
\end{minted}
\caption{Example of the required custom functions corresponding to 
the mean-field form in Eq.~\ref{eq:sign_free_afqmc}.}
\label{lst:custom}
\end{lstfloat}

Another example of a custom ansatz
is in recreating sign-free wave functions at half-filling for e.g. AFQMC, which use the form \cite{Qin2016}
\begin{equation}
\label{eq:sign_free_afqmc}
    \tilde{H}(M_i) = -t\sum_{\langle ij\rangle \sigma} c^\dagger_{i\sigma}c_{j\sigma} + h.c. + \sum_{i}M_i c^\dagger_{i\uparrow}c_{i\downarrow} + h.c.,
\end{equation}
with associated code shown in Listing~\ref{lst:custom}. 

To use the custom ansatz in an \autohf{} calculation, the \texttt{"ansatz"} setting must be set to \texttt{"CUSTOM"}
and we must provide the main function \texttt{solve\_hf} with \texttt{orbitalFunc} and \texttt{rdmFunc}.
We will also need to provide \texttt{solve\_hf} with a reference / initial $\alpha$.
Listing ~\ref{lst:run_custom_ansatz} demonstrates how to do this.
The ref state is randomly generated and additional noise 
is added individually to each parallel HF trial to give
each a different starting point.

\begin{figure}[h]
    \centering
    \includegraphics[width=0.5\textwidth]{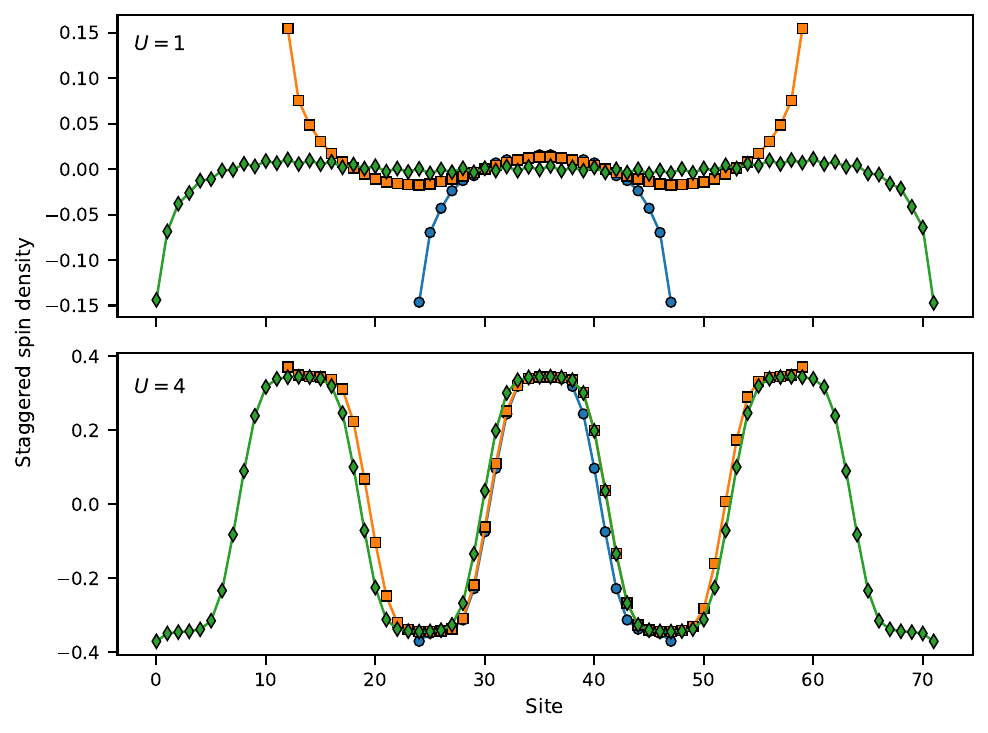}\includegraphics[width=0.5\textwidth]{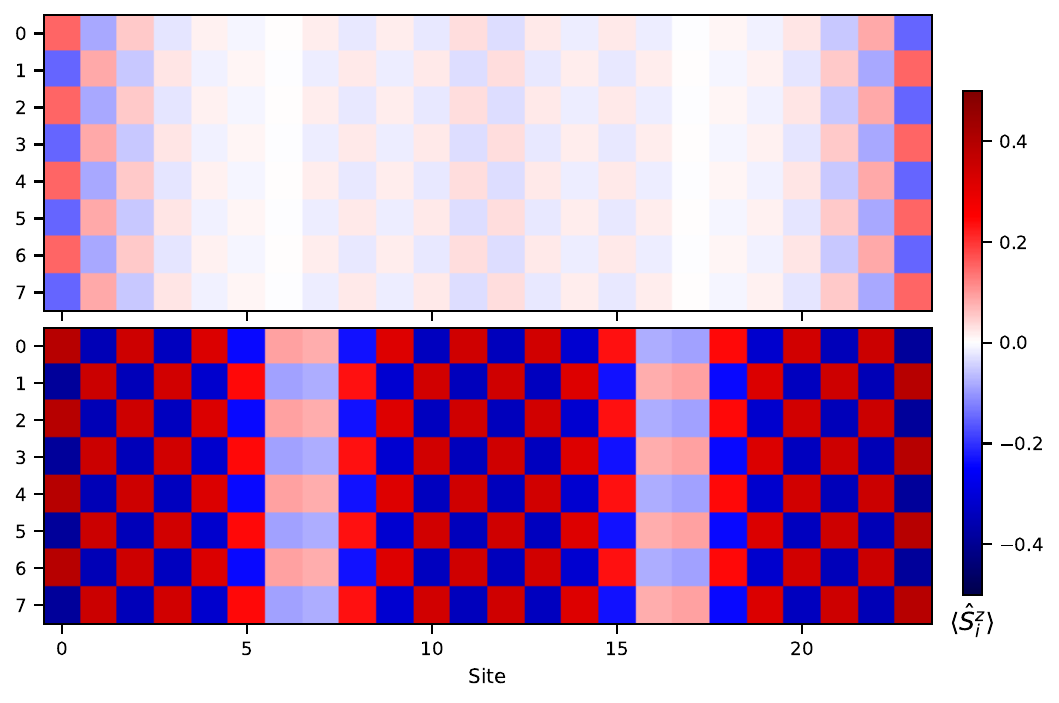}
    
    \caption{\textit{Left:} Measured staggered spin density $(-1)^{i_x+i_y}\langle \hat{S}^z_i\rangle $ using a UHF \autohf{} calculation of a Hubbard model with edge-pinning at $n=11/12$ i.e. $1/12$-doping), on a width-8 cylinder, with lengths $24$, $48$, and $72$. The top ($U=1$) shows the absence of long range order while the bottom ($U=4$) shows consistent order.
    \textit{Right:} Explicit spin density $\langle \hat{S}^z_i\rangle$ of the $24\times8$ cylinders. }
    \label{fig:stripes}
\end{figure}

\section{Examples}
\label{sec:examples}

The repository of \autohf{} holds nine sets of examples, ranging from basic setup to various custom enhancements and operating modes that a user may find useful.
Below we describe examples for calculations in Hubbard-like models and in quantum chemistry in two separate sections. Under the Hubbard model we show a number of systems and situations to help illustrate ways to use \autohf{}.

\subsection{Hubbard Models}
\label{ssec:Hubbard}

To illustrate the basic setup, consider a Hubbard model defined on some graph (e.g. square lattice with periodic boundary conditions) with hopping matrix $T$. 
The Hamiltonian is given by 
\begin{equation}
    H = T+V = \sum_{\langle ij\rangle \sigma} t_{ij}c^\dagger_{i\sigma}c_{j\sigma} + H.c. + U\sum_i n_{i\uparrow}n_{i\downarrow},
\end{equation}
where $T = t_{ij}$ is the graph of hopping terms and $U$ is uniform over all sites. The minimal script for this setup is shown in Listing~\ref{lst:basics}, which consists of setting up the Hamiltonian, determining the runtime parameters such as total number of steps or number of parallel batches, and passing those into wrappers for \autohf{} to use in its solver.

\begin{lstfloat}
\begin{minted}
[
frame=lines,
framesep=2mm,
baselinestretch=1.2,
fontsize=\footnotesize,
linenos
]{python}
# T, a (N,N) matrix defined elsewhere
steps = 100
batch_size = 4
nelec = (7,7)
U = 4

hf_settings = {
  "steps": steps,             # total steps
  "opt_method": "lbfgs",      # opt method
  "ansatz": "SD_ROT",         # orbital rotation ansatz
  "batch_size": batch_size,   # parallel batches
  "gpu": False,               # CPU vs GPU
  "nelec": nelec,             # number of electrons
  "noncollinear": False,      # UHF vs GHF form
}
# assumes T is the same for spin up and down
H = autohf.AutoHFHamiltonian(T=(T, T), U=U)
results = autohf.solve_hf(H, settings=hf_settings)
\end{minted}
\caption{Example of the most basic \autohf{} script, which runs a Hubbard model with hopping matrix $T$ and interaction strength $U$.}
\label{lst:basics}
\end{lstfloat}

For the following examples, the optimization of the orbitals is performed in two steps. First, a rotation matrix is optimized on top of a set of a reference state (\texttt{SD\_ROT} ) -- the non-interacting orbitals without pinning fields applied. This is done in parallel, optimizing several random initial parameters at once. States are then optimized to a loose tolerance (i.e. a tolerance value which is large) in order to obtain roughly the correct orders, while not necessarily being translationally invariant. 
After the top candidate states are found (we select the top 25\%), they are then optimized directly as  orbitals using the constraint dissolving (CD) variable change (\texttt{riemannian = True}) to a small tolerance, potentially removing the CD constraint for the last few steps.
Unless otherwise noted, all examples are optimized in this manner.

\subsubsection{Stripes in the Square Lattice Hubbard Model}

We show an example of UHF calculations in the simple Hubbard model on a square lattice. We will apply a pinning field, similar to that in Ref.~\cite{Xu2022}, of the form: $H_\text{pin} = h_\text{afm} (-1)^{i_x+i_y}\hat{S}^z_i$ for sites $i$ on the left or right edge. 
In principle there is no need to apply any pinning in UHF as the solution breaks translational symmetry anyway. However such a pinning is needed in a many-body calculation like AFQMC, if a broken symmetry solution is desired. In these situations, to produce an UHF trial wave function for AFQMC requires applying the same pinning field in the Hamiltonian.

Here, we use \autohf{} to obtain the correct UHF state without any prior assumptions of order or state. This may come at increased computational cost, but is useful for exploring new systems. 
The results of this optimization are shown in Fig.~\ref{fig:stripes}. 
As the interaction strength $U$ is increased, the system goes from having no long range spin density order to one with a long range staggered spin/charge density wave. An example of the explicit stripes of the spin density is shown on the right for both $U=1$ and $U=4$. 
While (unrestricted) Hartree-Fock calculations obtain the wrong phase boundary of these two states, the theory is able to capture the expected phases according to Ref.~\cite{Xu2022}. 

\subsubsection{Imposing symmetry}

\begin{figure}[h]
    \centering
    \includegraphics[width=0.6\textwidth]{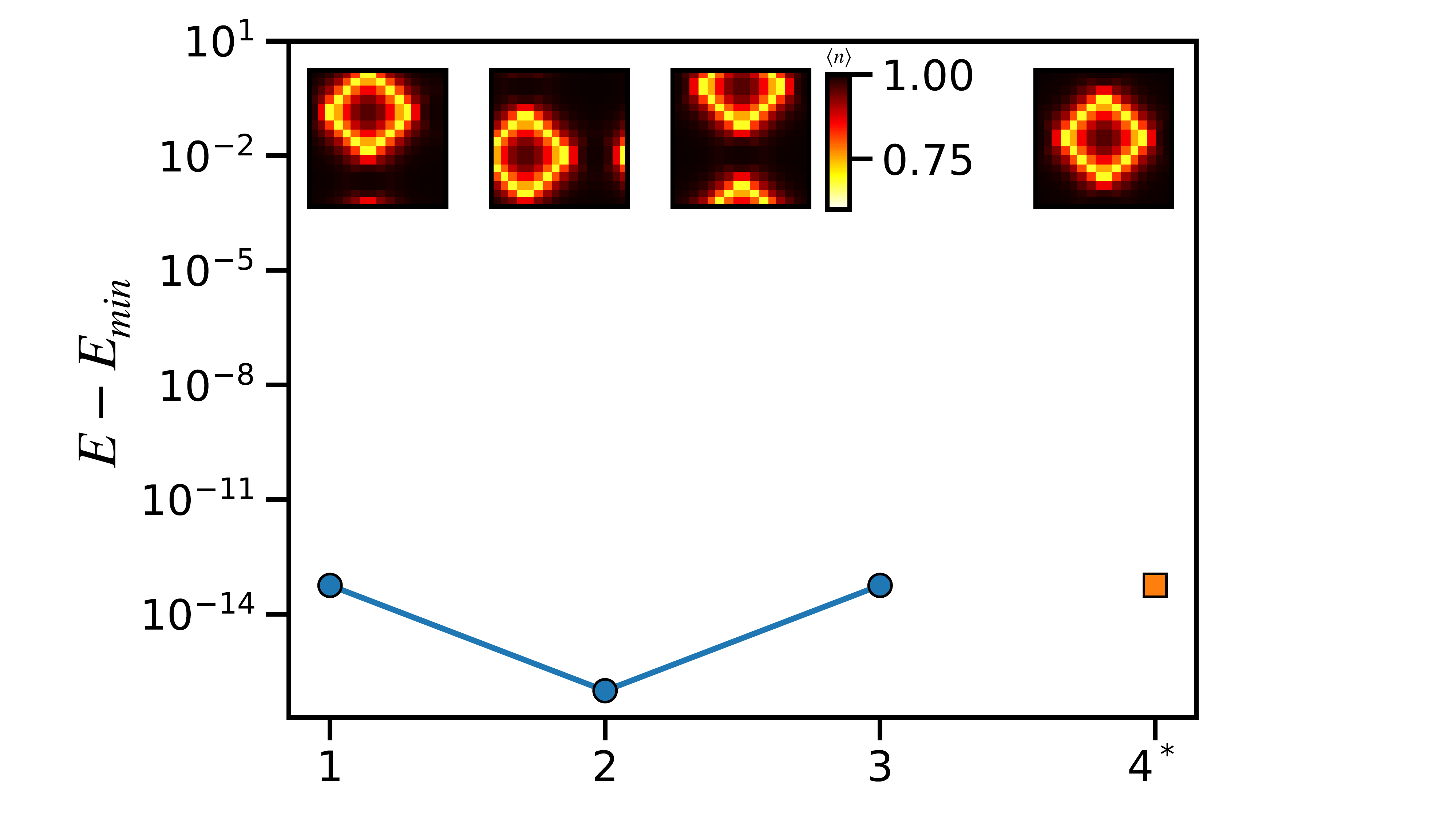}
    \caption{Energy differences from four different  random realizations of a square $16\times 16$ Hubbard model with 20 holes and $U=4$. The 4th seed was run with an additional energy term to penalize symmetry breaking. Inset: Charge density plots for each of the different realizations.}
    \label{fig:dope_hubb}
\end{figure}

While the previous example studied a Hamiltonian with a pinning term that breaks 
translational and rotational symmetries, here 
we consider the Hubbard model without pinning fields, in a square supercell, to illustrate how \autohf{} allows calculations to impose certain symmetries.
Using the same method as above, we compute four different random realizations of a $16 \times 16$ Hubbard model with 20 holes and $U=4$. Originally studied in Ref.\cite{Xu2011}, the authors used simulated annealing to carefully find the correct ground states of their mean field Hamiltonians. Here, we again provide no initial assumptions to \autohf{}, allowing the solution to automatically recover the symmetry properties. 

In Fig.~\ref{fig:dope_hubb}, the difference in energy of each of the four different initializations are shown. Each of the states are essentially degenerate up to small noise on the order of $O(10^{-12})$. 
The charge density $\langle n_{i} \rangle $ as a function of $i=(i_x,i_y)$ 
is plotted as a heat map in the 
inset for each of the realizations. 
Because of the translation invariance, there is a degenerate set of possible states which are discovered randomly. If one needs to  break this degeneracy and restore certain symmetry of the simulation supercell, \autohf{} allows for a custom energy penalty term to constrain the solution. We add an energy term 
\begin{equation}
 \lambda H_p = \lambda (||n_{(i_x,i_y)} - n_{(i_y,i_x)}||+||n_{(i_x,i_y)} - n_{(N_x-i_x,i_y)}||+||n_{(i_x,i_y)} - n_{(i_y,N_x-i_x)}||),
\end{equation}
which penalizes breaking of reflection symmetry about the diagonal ($x=y$), $y$, and $x$ axes of the simulation cell, shown in Listing~\ref{lst:custom_symm}. If $\lambda$ is not too large (which would select a completely uniform solution), the resulting state should be symmetrized. We set $\lambda=1$ and obtain a symmetrized state with essentially the same energy (when measured without the penalty term) as the other random realizations. Custom terms of this type are useful ways to target given symmetry or states in a flexible manner.

\begin{lstfloat}
\begin{minted}
[
frame=lines,
framesep=2mm,
baselinestretch=1.2,
fontsize=\footnotesize,
linenos
]{python}
lam = 1.0
Nx,Ny = 16,16
def lambda_Hp(state,rdms):
  r_uu,r_dd = rdms[0],rdms[1]
  density = (r_uu.diagonal()+r_dd.diagonal()).reshape(Nx,Ny)
  return lam*( jnp.linalg.norm(density-density.T)
            +jnp.linalg.norm(density-density[::-1])
            +jnp.linalg.norm(density-density[::-1].T))
\end{minted}
\caption{Symmetrized energy penalty term that allows selecting one of the translationally invariant realizations in Figure \ref{fig:dope_hubb}.}
\label{lst:custom_symm}
\end{lstfloat}

\subsubsection{Noncollinear Solution in Triangular Lattice Hubbard Model}

\begin{figure}[h]
    \centering
    \includegraphics[width=0.8\textwidth, trim={0 0.2cm 0 0},clip]{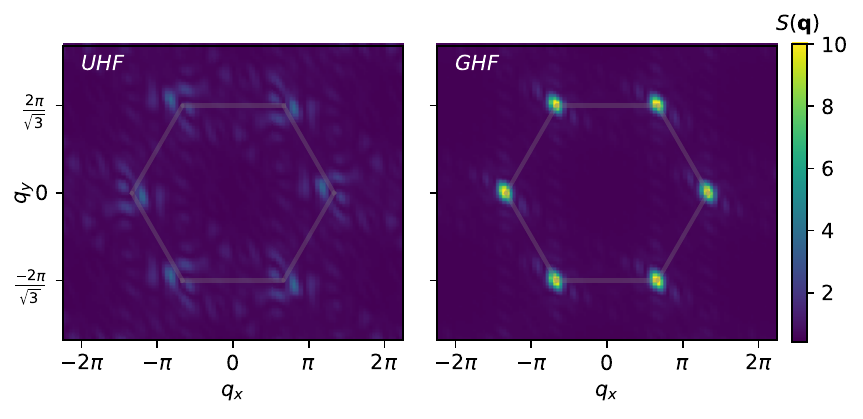}
    \caption{Measured static spin structure factor $S(\bm{q}) $ using a UHF \autohf{} (left) and GHF \autohf{} (right) calculation of a $12\times 12 $ $U=8$ Hubbard model on the triangular lattice at half-filling ($n=1$) with periodic boundaries. The UHF solution cannot properly form the 120$^\circ$ order and creates a frustrated order, while the GHF solution easily obtains the correct $120^\circ$ magnetic order.}
    \label{fig:tri_hubb}
\end{figure}

The choice between a collinear (UHF) and noncollinear (GHF) solution is crucial to determine the representation space of the mean field solution. We illustrate this for a strongly interacting ($U=8$) Hubbard model on the triangular lattice at half-filling ($n=1$). The mean-field ground state is  triangular, or $120^\circ$ AFM order, which is not within the possible solution space of UHF but is accessible with GHF.

To see this, we run a $12\times 12$ triangular Hubbard model at $U=8$ for both a UHF and GHF ansatz. After obtaining the lowest energy state for both a UHF and GHF representation we plot the static spin structure factor (Fig.~\ref{fig:tri_hubb}),
\begin{equation}
    S(\bm{q}) = \frac{1}{N}\sum_{i,j} e^{i\bm{q}\cdot (\bm{r}_i-\bm{r}_j)} \langle \bm{S}(\bm{r}_i)\cdot \bm{S}(\bm{r}_j)\rangle.
\end{equation}
The noncollinear state is approximately $6.6\%$ lower in relative energy and recovers rotational symmetry, while the collinear state forms an AFM pattern. Since the noncollinear representation can have an arbitrary total $\langle S^z\rangle$, we have used a penalty term (similar to the previous section) to obtain a state $\langle S^z\rangle \approx O(10^{-9})$. This calculation could be repeated with complex numbers to allow the spins to be outside  the x-z plane.

\subsection{Molecules}
\label{sec:ex_molecules}

\begin{figure}[h]
    \centering
    \includegraphics[width=0.45\textwidth]{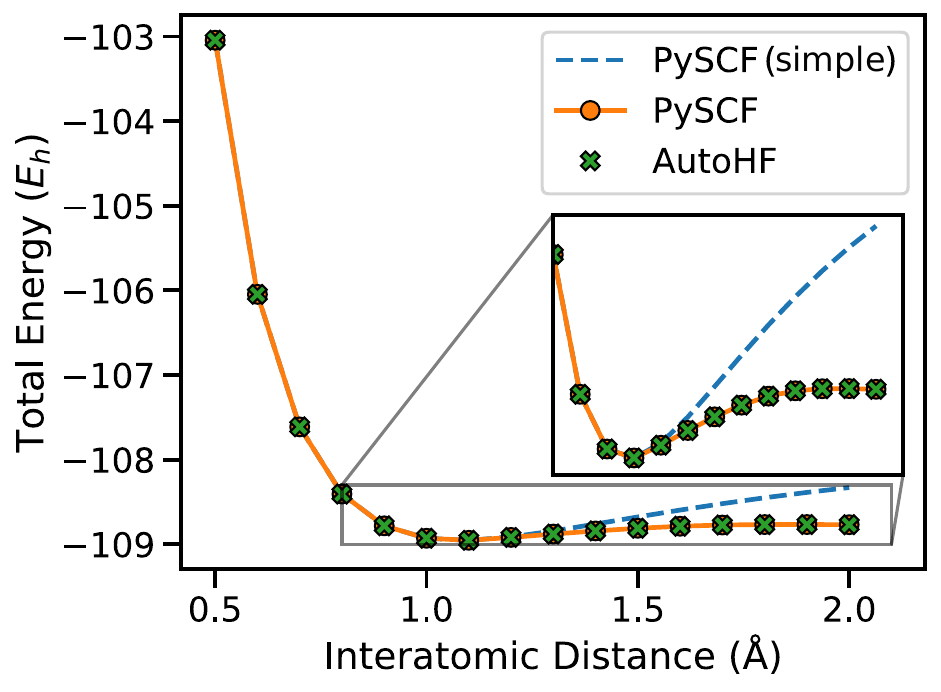}
    \caption{UHF energy for the bond stretching of N$_2$, using P\textsc{y}SCF (Naive), P\textsc{y}SCF with stability analysis, and \autohf{} for solutions. For \autohf, the Coulomb interaction is represented by a Cholesky decomposition. \textit{Inset}: zoomed in region showing the instability as the molecule dissociates.  }
    \label{fig:N2_diss}
\end{figure}

To highlight the versatility of \autohf, beyond the set of lattice-like Hamiltonian terms that are implemented by default,  we include an example of applying the same mean-field optimization philosophy to
quantum chemistry.

For such general systems, \autohf{} comes with the \texttt{autohf.terms.cholesky} submodule which allows 
a custom energy term based on a one-body Hamiltonian and an arbitrary Cholesky decomposition of the Coulomb interaction (Listing \Ref{lst:mol}).
We use P\textsc{y}SCF\cite{Sun2015,Sun2017,Sun2020} and 
 helper functions in the full example in the \autohf{} repository 
to generate that decomposition for an N\textsubscript{2} molecule in the cc-pVDZ Gaussian basis set at different bond lengths.
The N\textsubscript{2} molecule is a challenging case for HF at intermediate interatomic distances due to the near-degeneracy of many electon configurations. 
This is analagous to typical situations in lattice models.
For molecules, we find that the SD ansatz without the \texttt{"riemannian"} optimization converges reliably.
To speed up convergence, we initialize the \autohf{} calculation with the orbitals from the previous bond length but keep the remaining optimization procedure the same. 

P\textsc{y}SCF can robustly find the HF solution for molecules by first 
performing a simple HF calculation, which may converge to a higher-symmetry solution, and then performing a stability analysis to break appropriate symmetries
and find the correct minimum.
\autohf{}, on the other hand, will find the correct minimum by default and agrees with the more careful P\textsc{y}SCF calculation
as we demonstrate in Figure~\ref{fig:N2_diss}.

\begin{lstfloat}
\begin{minted}
[
frame=lines,
framesep=2mm,
baselinestretch=1.2,
fontsize=\footnotesize,
linenos
]{python}
# Defined elsewhere:
# onebody: (2, NMO, NMO)
# cholesky_operators: (Nchol, 2, NMO, NMO)

nuclear_repulsion = mol.energy_nuc()

def molecular_energy(state, rdms):
    return (
    autohf.terms.cholesky.hf_onebody(rdms, onebody)
      + autohf.terms.cholesky.hf_cholesky(rdms, cholesky_operators)
      + nuclear_repulsion
    )
\end{minted}
\caption{Custom energy for molecular calculations evaluating the expectation value of $H_\text{chol} = \sum_{\sigma ij} h_{\sigma ij} c^\dagger_{i\sigma} c_{j\sigma} - \frac 12 \sum_{n\sigma} \left(\sum_{ij} v_{n\sigma ij} c^\dagger_{i\sigma} c_{j\sigma}\right)^2 + E_\text{nuc-nuc}$ with the (shifted) \texttt{onebody} Hamiltonian $h_{\sigma ij}$ and the \texttt{cholesky\_operators} $v_{n\sigma ij}$. \texttt{Nchol} is the number of cholesky operators and \texttt{NMO} is the number of molecular orbitals.}
\label{lst:mol}
\end{lstfloat}

Finally, we apply \autohf{} to a set of 3d transition metal (TM) oxide diatomic molecules.
This class of molecules is known to pose a significant challenge to many-body approaches 
due to the presence of the 3d TM atoms as well as the non-trivial magnetic ground state of 
each molecule.
It is particularly challenging to converge HF for 3d TM oxide molecules due to many local minima
in the optimization landscape.
We again use the custom energy term in listing~\ref{lst:mol},
using \texttt{chol\_tol = 1e-6} here,
and the custom ROHF ansatz of listing~\ref{lst:custom_ansatz_rohf}.
Table~\ref{tbl:3dTMO} includes the HF energy of seven different 3d TM oxide diatomics
along with a reference HF value from the literature~\cite{3dTMO_benchmark}, the expected 
spin of the ground state, and the measured value of 
$\langle \hat{S}^2 \rangle$ in \autohf.
The \autohf{} HF energy agrees well with the results from the literature
with all discrepancies on the order of the Cholesky tolerance, $1\times10^{-6}$, 
with the exception of the VO molecule.
In this case, \autohf{} found a lower variational energy, and therefore better HF solution, than that found in this particular reference.
We verified this by using the \autohf{} HF solution as the initial state in a P\textsc{y}SCF
 HF calculation. 
We found that PySCF immediately converged with the same
 variational energy as the \autohf{} calculation.

\begin{table}[h]
\centering
\begin{tabular}{lll}
  \hline
  \textbf{Molecule} & \autohf{} $E_\text{HF}$ & \textbf{Reference} $E_\text{HF}$~\cite{3dTMO_benchmark} \\
  \hline
    ScO & $\mathbf{-61.829343}$ & $-61.829343$ \\
    TiO & $-73.269736$ & $\mathbf{-73.269736}$ \\
    VO  & $\mathbf{-86.400549}$ & $-86.400212$ \\
    CrO & $-101.838135$ & $\mathbf{-101.838135}$ \\
    MnO & $-119.133239$ & $\mathbf{-119.133240}$ \\
    FeO & $\mathbf{-138.660099}$ & $-138.660097$ \\
    CuO & $\mathbf{-212.238840}$ & $-212.238830$ \\
  \hline
\end{tabular}
\caption{HF energies for the 3d TM oxide molecules computed using \autohf. Bold indicates the lower energy. }
\label{tbl:3dTMO}
\end{table}

\section{Conclusion}
\label{sec:conclusion}

The \autohf{} package provides a flexible implementation of direct variational Hartree-Fock optimization that is designed to support rapid experimentation with different models and Hamiltonians. 
Implemented entirely in Python using JAX, \autohf{} takes advantage of automatic differentiation 
allowing Hartree-Fock calculations to be defined, controlled, and extended directly from user scripts.
\autohf{} includes several standard, predefined ansätze, with support for user-defined custom ansätze, which can
all be directly applied in HF calculations for a range of composable interacting Hamiltonians. %
Since \autohf{} is formulated in second quantization, it is agnostic to the graph on which the interacting Hamiltonian is defined and can, therefore, be applied to arbitrary lattices and dimensions.
In addition to the capabilities already built in, \autohf{} presents a platform on which extensions, generalizations, and new 
features can be conveniently introduced or prototyped.

A central feature of \autohf{} is the ability to include custom Hamiltonian terms and energy penalties directly in the variational objective, which provides fine-grained control over the properties of the resulting Hartree-Fock solution. 
In particular, quadratic penalty terms that vanish at the minimum allow constraints to be imposed without affecting the final energy, making it possible to select specific solutions from nearly degenerate manifolds.
We demonstrated these capabilities on a range of representative lattice and ab initio systems.

The examples presented here showcase a variety of techniques and robustness in the methodology. For the Hubbard model on a long cylinder of size $72 \times 8$, \autohf{} converged to a stripe-ordered solution with the correct wavelength without requiring a symmetry-breaking initial guess, in contrast to the strong dependence on initialization that is often observed in traditional SCF implementations.
On a $16 \times 16$ Hubbard lattice, a custom energy penalty was used to suppress symmetry-broken solutions, allowing \autohf{} to select a symmetric state from a set of nearly degenerate Hartree-Fock solutions separated by energies on the order of $10^{-12}t$.
We also demonstrated the ease with which different HF variants can be explored using the triangular lattice Hubbard model, where GHF naturally produced the expected $120^\circ$ magnetic order.
In ab initio quantum chemistry
\autohf{} was able to find the global minimum with no need to perform a stability analysis in N\textsubscript{2} bondbreaking and in a set of $3d$ transition metal oxide molecules.
Taken together, these results show that \autohf{} 
is able to find global minima with remarkable reliability, providing
is a robust and extensible platform for generating HF states with controlled symmetry and qualitative features.

\autohf{} provides a flexible foundation for a wide range of many-body applications.
For example, more species (SU(3), SU(4), etc) and more general forms of the Hamiltonian (e.g., valley or flavor degrees of freedom) are straightforward to include.
While \autohf{} is 
designed for single Slater determinant ansätze, 
it naturally supports more general constructions through user-defined ansätze, including multi-determinant forms.
Future extensions may include more general
ansätze that go beyond Slater determinants
such as Hartree-Fock--Bogoliubov forms~\cite{Shi2017-HFB,Vitali2024-HFB}, while retaining the same 1-rdm variational and automatic differentiation-based optimization framework.
It would also be interesting to explore extensions to finite temperature HF~\cite{Scholle2023} in the framework presented here.

Finally, we comment that \autohf{} was designed to handle broad classes of Hamiltonians, emphasizing generality, flexibility, and convenience.
Absolute speed was less of
a focus, especially in traditional domains of applications such as electronic structure and quantum chemistry, where decades of development has gone into SCF optimizations.
In the examples we tested, \autohf{} often showed surprising efficiency in finding the global minimum. 
It will be interesting to perform more systematic comparisons with SCF. 
With more tailored efforts and, perhaps more importantly, with the rapid pace of advances in variational approaches driven by 
machine-learning developments,  
\autohf{} can be expected to become a valuable alternative for ab initio calculations in solids and molecules as well. 

\section*{Acknowledgements}
The authors thank Chris Roth, Leonardo dos Anjos Cunha, and Yubo Yang for helpful discussions and initial testing. The Flatiron Institute is a division of the Simons Foundation. 

\clearpage
\begin{appendix}
\numberwithin{equation}{section}
\section{Advanced \autohf{} Settings}
\label{appendix:autohf_settings}
In this section, we describe the advanced options in \autohf.

Table~\ref{atbl:advanced_settings} lists all of the advanced options along with the default value,
and a description of the setting.

\vspace{0.5cm}
\begin{threeparttable}
\begin{tabular}{ llp{8cm} }
 \hline
  \textbf{Parameter} & \textbf{Default} & \textbf{Description} \\
 \hline
  approx\_expm & \texttt{False} & If \texttt{True}, use a fast, but approximate, of matrix exponentiation. \\
  dump\_batch & \texttt{False} & Only applies when ``output'' is set. Dump the orbitals from all trials to the output file, as opposed to only those of the best HF solution. \\
  eta & 0.1 & the step size for gradient decent. Has no effect for "lbfgs" or "basin" optimization methods. \\
  force\_complex & \texttt{False} & If \texttt{True}, forces the HF orbitals to be complex-valued.  \\
    output & \texttt{None} &  Name of file to save results to. The format and data saved is determined from the file name. \texttt{"*.npy"} will save the orbitals to the numpy binary format. \texttt{"*.h5"} will save $\alpha$ for the best HF solution and for all trials, energies for the best HF solution and for all trials, the orbitals of the best variational state, the \autohf{} settings, the reference Slater determinant, the number of electrons used in an HDF5 file. \texttt{".pkl"} or \texttt{".pickle"} will save the same information as the HDF5 format, but as a Python pickle file. HDF5 is recommended over pickle files for security reasons\tnote{1}.
\\
  riemannian & \texttt{False} & If \texttt{True}, use the constraint dissolving approach to impose optimization on a Stiefel manifold. Only applicable to \texttt{SD} or \texttt{SDC} ansätze. \\
  state0\_scale & 0.01 & used to control the amount of random noise introduced to the initial $\alpha$. The details of how noise is applied depends on the specific ansatz used. \\
  verbose & \texttt{False} & if \texttt{True}, turns on verbose debug printing. This will significantly slow down calculations. \\
  jaxoptargs & \texttt{None} & a dictionary of keyword settings to pass along to the jaxopt LBFGS optimizer. We refer you to the \href{https://jaxopt.github.io/stable/\_autosummary/jaxopt.LBFGS.html}{jaxopt documentation} for more. \\
  \hline
\end{tabular}
\begin{tablenotes}
     \item[1] Only use Python pickle files that you trust. Arbitrary code can be encoded and then executed when loading a pickle file.
\end{tablenotes}
\caption{Extended reference for the entries of the \texttt{settings} dictionary passed to the \texttt{solve\_hf} function.}
\label{atbl:advanced_settings}
\end{threeparttable}

\end{appendix}

\bibliography{main.bib}

\end{document}